\documentclass[preprint,showpacs,preprintnumbers]{revtex4}
\usepackage{amsmath}
\usepackage{graphicx}
\usepackage{dcolumn}
\usepackage{bm}
\usepackage{epsfig}
\usepackage[T1]{fontenc}
\usepackage{ae,aecompl}
\usepackage{appendix}

\begin{document}

\title{Lattice Boltzmann study on Kelvin-Helmholtz instability: the roles of
velocity and density gradients}
\author{Yanbiao Gan$^{1,2,3}$, Aiguo Xu$^2$\footnote{
Corresponding author. E-mail: Xu\_Aiguo@iapcm.ac.cn}, Guangcai
Zhang$^2$, Yingjun Li$^1$} \affiliation{1, State Key Laboratory for
GeoMechanics and Deep Underground Engineering, SMCE, China
University of Mining and Technology (Beijing), Beijing 100083,
P.R.China\\
2, National Key Laboratory of Computational Physics, \\
Institute of Applied Physics and Computational Mathematics, P. O.
Box 8009-26, Beijing 100088, P.R.China \\
3, North China Institute of Aerospace Engineering, Langfang 065000,
P.R.China }
\date{\today }

\begin{abstract}
A two-dimensional lattice Boltzmann model with 19 discrete
velocities for compressible fluids is proposed. The fifth-order
Weighted Essentially Non-Oscillatory (5th-WENO) finite difference
scheme is employed to calculate the convection term of the lattice
Boltzmann equation. The validity of the model is verified by
comparing simulation results of the Sod shock tube with its
corresponding analytical solutions. The velocity and density
gradient effects on the Kelvin-Helmholtz instability (KHI) are
investigated using the proposed model. Sharp density contours are
obtained in our simulations. It is found that, the linear growth
rate $\gamma$ for the KHI decreases with increasing the width of
velocity transition layer ${D_{v}}$ but increases with increasing
the width of density transition layer ${D_{\rho}}$. After the
initial transient period and before the vortex has been well formed,
the linear growth rates, $\gamma_v$ and $\gamma_{\rho}$, vary with
${D_{v}}$ and ${D_{\rho}}$ approximately in the following way, $%
\ln\gamma_{v}=a-bD_{v}$ and $\gamma_{\rho}=c+e\ln D_{\rho} ({D_{\rho}}<{%
D_{\rho}^{E}})$, where $a$, $b$, $c$ and $e$ are fitting parameters, and ${%
D_{\rho}^{E}}$ is the effective interaction width of density
transition layer. When ${D_{\rho}}>{D_{\rho}^{E}}$ the linear growth rate $%
\gamma_{\rho} $ does not vary significantly any more. One can use
the hybrid effects of velocity and density transition layers to
stabilize the KHI. Our numerical simulation results are in general
agreement with the analytical results [L. F. Wang, \emph{et al.},
Phys. Plasma \textbf{17}, 042103 (2010)].
\end{abstract}

\preprint{} \maketitle

\section{Introduction}

As a mesoscopic approach and a bridge between the molecular dynamics
method at the microscopic level and the conventional numerical
method at the macroscopic level, the lattice Boltzmann (LB) method
\cite{Succi-Book} has been successfully applied to various fields
during the past two decades, ranging from the multiphase system
\cite{Yeomans,XGL1,Sofonea-multiphase}, magnetohydrodynamics
\cite{Chensy-PRL-1991,Succi-PRA-1991,CICP-2008}, reaction-diffusion
system \cite{reactive-1,reactive-2,reactive-3}, compressible fluid
dynamics \cite{Sunch-PRE-1998,Katato,Watari,Xu-compressible,Xu-compressible-2,Xu-compressible-3,QuKun,HYL}%
, simulations of linear and nonlinear partial differential equations
\cite{PDE-1}, etc. Its popularity is mainly owed to its kinetic
nature \cite{kinetic nature}, which makes the physics at mesoscopic
scale can be incorporated easily. As a result, this approach
contains more physical connotation than Navier-Stokes or Euler
equations based on the continuum hypothesis. Its popularity is also
owed to its linear convection term, easy implementation of boundary
conditions, simple programming and high efficiency in parallel
computing, etc.

The Kelvin-Helmholtz instability (KHI) occurs when two fluids flow
with different tangential velocities \cite{KH-BOOK}. Under the
condition of the KHI, small perturbations along the interface
between two fluids undergo linear
\cite{PFA-1993,PF-1982,PF-1980,wang-pop-2009,wang-pop-2010} and
nonlinear growth stages
\cite{PRL-1999,PRL-2002,PRE-2004,POP-2005,wang-epl-2009}, and may
evolve into turbulent mixing through nonlinear interactions. The KHI
has attracted much attention because of its significance in both
fundamental research and engineering applications \cite{HEDP}. On
the one hand, in fundamental research, KHI is of great importance in
the fields of turbulent mixing \cite{turbulent}, supernovae dynamics
\cite{supernovae-1,supernovae-2,supernovae-3}, and the interaction
of the solar wind with the earth's magnetosphere \cite{earth
magnet}, etc. On the other hand, in engineering applications, the
KHI plays a key role in small-scale mixing of Rayleigh-Taylor (RT)
\cite{pop-2010-2,epl-2010-2} and Richtmyer-Meshkov (RM)
instabilities in inertial confinement fusion (ICF)
\cite{ICF-3,ICF-4,ICF-book}. In the final regime of RT and RM
instabilities, KHI is initiated due to the shear velocity difference
at the spike tips \cite{KH-RM-RT}, and, therefore, the appearance of
the KHI aggravates the development of final nonlinearity of RT
instability or RM instability, and quickens the process of fluid
flock mixing round the interface.

In the present work, we propose a two-dimensional LB mode to
simulate the KHI. The model consists of 19 discrete velocities in
six directions and allows to recover the compressible Euler
equations in the continuum limit. Actually, this phenomenon has been
studied extensively by many researchers using experimental
measurements, theoretical analysis and more recently by numerical
simulations during the past decades
\cite{KH-BOOK,wang-pop-2009,wang-pop-2010,wang-epl-2009,HEDP,turbulent}.
Those results indicate that the evolution of KHI depends on the
density ratio, viscosities, surface tension, compressibility and
others. Although the basic behavior of the KHI has been studied
extensively, to the best of our knowledge, the use of the LB method
to investigate the evolution of this phenomenon is still very
limited. In this paper, with the LB method we focus on the velocity
and the density gradient effects on KHI. The rest of the paper is
organized as follows. In Section II, using the Chapman-Enskog
analysis, we show that the current model can recover the Euler
equations in the continuum limit. The numerical scheme and the
validation of the model will be performed in Section III. In Section
IV, we show numerical simulation results on the KHI for various
widths of velocity and density transition layers. Both the two kinds
of gradients effects are investigated carefully. Finally, in Section
V conclusions and discussions are drawn.

\section{Formulation of the LB model}

The model described here is inspired by the previous work of Watari and
Tsutahara \cite{Watari}, which is based on a multispeed approach and where
the truncated local equilibrium distribution uses global coefficients.

\subsection{Designing of the discrete velocity model}

In the present study, we will propose a finite difference LB (FDLB)
model for compressible flows. Within this scheme, the physical
symmetry of hydrodynamic systems can be more conveniently recovered
by the use of discrete velocity model with high spatial isotropy.
For completeness, let us start with the standard LB equation, an
exact lattice-based dynamical equation expressed in terms of
particle streaming and the Bhatnagar-Gross-Krook (BGK) collision
operator
\begin{equation}
f_{i}^{'}(\mathbf{r}^{'}+\mathbf{v}_{i}^{'}\Delta t^{'},t^{'}+\Delta
t^{'})
 -f_{i}^{'}(\mathbf{r}^{'},t^{'})=-\frac{1}{\tau^{'}/\Delta t^{'}}(f_{i}^{'}-f_{i}^{(0)'})\text{,}
\end{equation}
where $f_{i}^{'}$ is the distribution function along the $i$
direction, $f_{i}^{(0)'}$ is its corresponding equilibrium
distribution function, $\mathbf{r}^{'}$ is the spatial coordinate,
$\mathbf{v}_{i}^{'}$ is the particle velocity in the $i$ direction,
$\Delta t^{'}$ is the time step, and $\tau^{'}$ is relaxation time
determining the viscosity of fluid. For the standard LB method, all
the particle velocities are restricted to those that exactly link
the lattice sites in unit time. In other words, there is a banding
of discretizations of space and time, which is an inconvenience in
the use of the standard LB model to study the thermohydrodynamic
problems.

The standard LB equation can be approximately expressed as the
discrete velocity Boltzmann equation
\begin{equation}
\frac{\partial f_{ki}^{'}}{\partial {{t}^{'}}}+\mathbf{v}%
_{ki}^{'}\cdot \frac{\partial f_{ki}^{'}}{\partial {{%
\mathbf{r}}^{'}}}=-\frac{1}{{{\tau }^{'}}}%
(f_{ki}^{'}-f_{ki}^{(0)'}) \text{,} \label{BGK-Eq}
\end{equation}
where $\mathbf{v}_{ki}^{'}$ represents the discrete velocity model,
subscript $k$ indicates the $k$th group of the particle velocities
whose speed is $v_{k}^{'}$. Eq.(\ref{BGK-Eq}) may be written in a
dimensionless form by choosing
three independent reference variables, the characteristic flow length scale $L_{0}$%
, the reference speed $c_{s}$ and the reference density $\rho _{0}$.
The characteristic time $t_{0}$ can be expressed as
$t_{0}=L_{0}/c_{s}$. With these reference variables, we have the
following dimensionless variables
\begin{equation}
t=\frac{{{t}^{^{\prime }}}}{{{t}_{0}}}\text{, }r=\frac{{{r}^{^{\prime }}}}{{{%
r}_{0}}},\text{ }{{v}_{ki}}=\frac{v_{ki}^{^{\prime }}}{{{c}_{s}}}\text{, }{{f%
}_{ki}}=\frac{f_{ki}^{^{\prime }}}{{{\rho }_{0}}},\text{ }\tau
=\frac{{{\tau }^{^{\prime }}}}{{{t}_{0}}}\text{.}
\end{equation}%
Consequently, we have the following dimensionless lattice Boltzmann
equation
\begin{equation}
\frac{{\partial f_{ki}}}{{\partial t}}+\mathbf{v}_{ki}.\frac{\partial }{{%
\partial \mathbf{r}}}f_{ki}=-\frac{{1}}{\tau }(f_{ki}-f_{ki}^{{(0)}})\text{.}
\label{BGK-Eq-nonD}
\end{equation}%

To formulate a FDLB model, the next step is to chose an appropriate
dimensionless discrete velocity model $\mathbf{v}_{ki}$. For a
discrete velocity model described by
\begin{equation}
\mathbf{v}_{0}=0\text{, }\mathbf{v}_{ki}=v_{k}[\cos (\frac{{i2\pi}}{N})\text{%
,}\sin (\frac{{i2\pi}}{N})]\text{,}  \label{DVM}
\end{equation}%
the $n$th rank tensor is defined as
\begin{equation}
T_{\alpha_{1}\alpha _{2}...\alpha _{n}}^{(n)}=\sum\limits_{i=1}^{n}{%
v_{i\alpha_{1}}}v_{i\alpha_{2}}...v_{i\alpha_{n}}\text{,}  \label{Tensor}
\end{equation}%
where $\alpha_{1}$, $\alpha_{2}$,...,$\alpha _{n}$ indicate either
the $x $ or $y$ component. The tensor is isotropic if it is
invariant under the coordinate rotation and the reflection. As for
being isotropic, the odd rank tensors should be zero and the even
rank tensors should be the sum of all possible products of Kronecker
delta. In this study, the discrete velocity model with $N=6$ is used
to construct a FDLB model. For this discrete velocity model it is
easy find that the odd rank tensors are zero, and the even rank
tensors generally have the following forms\qquad
\begin{equation}
\sum\limits_{i{=1}}^{{6}}{v_{ki\alpha }}v_{ki\beta
}=3v_{k}^{2}\delta_{\alpha\beta}\text{,}  \label{moment-1}
\end{equation}%
\begin{equation}
\sum\limits_{i{=1}}^{{6}}{v_{ki\alpha }}v_{ki\beta }v_{ki\gamma }v_{ki\chi }=%
\frac{{3}}{{4}}v_{k}^{4}\Delta _{\alpha \beta \gamma \chi }\text{,}
\label{moment-2}
\end{equation}%
where
\begin{equation}
\Delta_{\alpha\beta\gamma\chi}{=}\delta_{\alpha\beta}\delta_{\gamma
\chi}+\delta_{\alpha\gamma}\delta_{\beta\chi}+\delta_{\alpha\chi
}\delta_{\beta\gamma}\text{.}
\end{equation}%
Therefore, this discrete velocity model is isotropic up to, at
least, its fifth rank tensor.

\subsection{Recovering of the Euler equations}

The macroscopic density $\rho $, momentum $\rho \mathbf{u}$ and temperature $%
T$ are calculated as the moments of the local equilibrium distribution
function
\begin{equation}
\sum\limits_{ki}{f_{ki}^{(0)}}\left(
\begin{array}{l}
1 \\
\mathbf{v}_{ki} \\
\frac{1}{2}(\mathbf{v}_{ki}-\mathbf{u})^{2}%
\end{array}%
\right) =\left(
\begin{array}{l}
\rho \\
\rho \mathbf{u} \\
\rho T=P%
\end{array}%
\right) \text{.}  \label{n-u-T}
\end{equation}

Besides the moments in Eq.(\ref{n-u-T}), Chapman-Enskog analysis indicates
the following additional ones are needed to satisfy in order to recover the
hydrodynamic equations at Euler level
\begin{equation}
\sum\limits_{ki}{v_{ki\alpha }v_{ki\beta}}f_{ki}^{{(0)}}=P\delta
_{\alpha\beta}+\rho u_{\alpha}u_{\beta}\text{,}  \label{moment-4}
\end{equation}%
\begin{equation}
\sum\limits_{ki}{\frac{{1}}{{2}}v_{ki}^{{2}}}v_{ki\alpha}f_{ki}^{{(0)}%
}=u_{\alpha }({2}P+\frac{{1}}{{2}}\rho u^{{2}})\text{.}  \label{moment-5}
\end{equation}

To perform the Chapman-Enskog expansion on the two sides of Eq.(\ref{BGK-Eq-nonD}%
), we introduce expansions
\begin{equation}
f_{ki}=f_{ki}^{(0)}+\epsilon f_{ki}^{(1)}+\epsilon
^{2}f_{ki}^{(2)}+\cdot\cdot\cdot \text{,}  \label{CE-1}
\end{equation}%
\begin{equation}
\frac{\partial }{{\partial t}}=\epsilon \frac{\partial }{{\partial t_{0}}}%
+\epsilon ^{2}\frac{\partial }{{\partial t_{1}}}+\cdot \cdot \cdot \text{,}
\label{CE-2}
\end{equation}%
\begin{equation}
\frac{\partial}{{\partial r_{\alpha }}}=\epsilon\frac{\partial }{{\partial
r_{1\alpha }}}\text{,}  \label{CE-3}
\end{equation}%
where $\epsilon\ll 1$ is the Knudsen number. The introducing of $\epsilon $
is equivalent to stipulating that the gas is dominated by large collision
frequency. The $f_{ki}^{(0)}$ is the zeroth order, $f_{ki}^{(1)}$, $\frac{%
\partial }{{\partial t_{0}}}$ and $\frac{\partial}{{\partial r_{1\alpha}}}$
are the first order, $f_{ki}^{(2)}$ and $\frac{\partial }{{\partial
t_{1}}}$ are the second order terms of the Knudsen number $\epsilon
$. Substituting Eqs.(\ref{CE-1})-(\ref{CE-3}) into
Eq.(\ref{BGK-Eq-nonD}) and comparing the coefficients of the same
order of $\epsilon$ give
\begin{equation}
\epsilon ^{1}:\frac{{\partial f_{ki}^{(0)}}}{{\partial t_{0}}}+\frac{\partial%
}{{\partial r_{1\alpha}}}(f_{ki}^{(0)}v_{ki\alpha})=-\frac{1}{\tau }f_{ki}^{{%
(1)}}\text{,}  \label{1st}
\end{equation}%
\begin{equation}
\epsilon ^{2}:\frac{{\partial f_{ki}^{(0)}}}{{\partial t_{1}}}+\frac{{%
\partial f_{ki}^{(1)}}}{{\partial t_{0}}}+\frac{\partial}{{\partial
r_{1\alpha }}}(f_{ki}^{(1)}v_{ki\alpha})=-\frac{1}{\tau}f_{ki}^{{(2)}}\text{,%
}  \label{2nd}
\end{equation}%
\begin{equation}
\epsilon ^{j}:\frac{{\partial f_{ki}^{(0)}}}{{\partial t_{j-1}}}+\frac{{%
\partial f_{ki}^{(1)}}}{{\partial t_{j-2}}}+...\frac{{\partial f_{ki}^{(j-1)}%
}}{{\partial t_{0}}}+\frac{\partial }{{\partial r_{1\alpha }}}%
(f_{ki}^{(j-1)}v_{ki\alpha })=-\frac{1}{\tau }f_{ki}^{{(j)}}\text{.}
\label{jth}
\end{equation}%
Summing Eq.(\ref{1st}) over $k,i$ gives
\begin{equation}
\frac{{\partial}\rho}{{\partial t_{0}}}+\frac{\partial}{{\partial r_{1\alpha}%
}}(\rho u_{\alpha })=0\text{.}
\end{equation}%
Summing Eq.(\ref{jth}) over $k,i$ gives
\begin{equation}
\frac{{\partial}\rho}{{\partial t_{j-1}}}=0\mathrm{,}{(}j\geq 2{)}\text{.}
\end{equation}%
Using Eq.(\ref{CE-2}) and Eq.(\ref{CE-3}) gives
\begin{equation}
\frac{{\partial}\rho}{{\partial t}}+\frac{\partial}{{\partial r_{\alpha}}}%
(\rho u_{\alpha})=0\text{.}
\end{equation}%
It is clear that the continuity equation can be derived at any order of $%
\epsilon $. Performing the operator $\sum\limits_{ki}{v_{ki\alpha }}$ to the
two sides of Eq.(\ref{1st}) gives
\begin{equation}
\frac{\partial}{{\partial t}_{0}}(\rho u_{\alpha})+\frac{\partial }{{%
\partial r_{1\beta}}}[\rho (T\delta_{\alpha\beta}+u_{\alpha }u_{\beta })]=0%
\text{.}
\end{equation}
Performing the operator $\sum\limits_{ki}{v_{ki\alpha}}$ to the two sides of
Eq.(\ref{2nd}) gives
\begin{equation}
\frac{\partial }{{\partial t}_{1}}(\rho u_{\alpha })=0\text{.}
\end{equation}
Using Eq.(\ref{CE-2}) and Eq.(\ref{CE-3}) gives the moment equation at the
Euler level
\begin{equation}
\frac{\partial}{{\partial t}}(\rho u_{\alpha })+\frac{\partial }{{\partial
r_{\beta }}}[\rho (T\delta_{\alpha\beta}+u_{\alpha }u_{\beta })]=0\text{.}
\end{equation}
Performing operator $\sum\limits_{ki}{v_{ki}^{{2}}/2}$ to both sides of Eq.(%
\ref{1st}) gives
\begin{equation}
\frac{\partial }{{\partial t}_{0}}[n(T+\frac{{u^{2}}}{2})]+\frac{\partial }{{%
\partial r_{1\beta }}}[\rho u_{\beta }(2T+\frac{{u^{2}}}{2})]=0\text{.}
\end{equation}
Performing operator $\sum\limits_{ki}{v_{ki}^{{2}}/2}$ to both sides of Eq.(%
\ref{2nd}) gives
\begin{equation}
\frac{\partial }{{\partial t}_{1}}[n(T+\frac{{u^{2}}}{2})]=0\text{.}
\end{equation}
Using Eq.(\ref{CE-2}) and Eq.(\ref{CE-3}) gives the energy equation at the
Euler level
\begin{equation}
\frac{\partial }{{\partial t}}[n(T+\frac{{u^{2}}}{2})]+\frac{\partial }{{%
\partial r_{\beta }}}[\rho u_{\beta }(2T+\frac{{u^{2}}}{2})]=0\text{.}
\end{equation}

\subsection{Formulation of the discrete equilibrium distribution function}

We now formulat the discrete equilibrium distribution function based
on the discrete velocity model with $N=6$. The requirement
Eq.(\ref{moment-5}) contains up to the third order of the flow
velocity $u$. It is reasonable to expand the local equilibrium
distribution $f_{ki}^{{(0)}}$ in polynomial of the flow velocity up
to the third order
\begin{eqnarray}
f_{ki}^{(0)} &=&\frac{\rho }{{2\pi T}}\exp [-\frac{{v_{k}^{2}}}{{2T}}]\exp [-%
\frac{{u^{2}-2uv_{k}}}{{2T}}]  \notag \\
&=&\rho F_{k}[(1-\frac{{u^{2}}}{{2T}})+\frac{{v_{ki\varepsilon
}u_{\varepsilon }}}{T}(1-\frac{{u^{2}}}{{2T}})+\frac{{v_{ki\varepsilon
}v_{ki\eta }u_{\varepsilon }u_{\eta }}}{{2T^{2}}}  \notag \\
&&+\frac{{v_{ki\varepsilon }v_{ki\eta }v_{ki\xi }u_{\varepsilon }u_{\eta
}u_{\xi }}}{{6T^{3}}}]+\cdot \cdot \cdot \text{,}
\end{eqnarray}%
where
\begin{equation}
F_{k}=\frac{1}{{2\pi T}}\exp [-\frac{{v_{k}^{2}}}{{2T}}]  \label{FFKK}
\end{equation}%
is a function of temperature $T$ and particle velocity ${v_{k}}$.
The truncated equilibrium distribution function $f_{ki}^{(0)}$
contains the third rank tensor of the particle velocity and the
requirement Eq.(\ref{moment-4}) contains the second rank tensor.
Thus, the discrete velocity model should have an isotropy up to its
fifth rank tensor. So the discrete velocity model with $N=6$ is an
appropriate choice.

To numerically calculate the equilibrium distribution function, one
needs first calculate the global factor $F_{k}$. It should be noted
that $F_{k}$ can not be calculated directly from its definition
Eq.(\ref{FFKK}). It should take values in such a way that satisfies
Eqs.(\ref{n-u-T})-(\ref{moment-5}).

To satisfy the first equation in Eq.(\ref{n-u-T}), we require
\begin{equation}
\sum\limits_{ki}{F_{k}}=1\text{,}  \label{FK-1}
\end{equation}%
\begin{equation}
\sum\limits_{ki}{F_{k}v_{ki\varepsilon }v_{ki\eta }u_{\varepsilon }u_{\eta }}%
=Tu^{2}\text{.}  \label{FK-2}
\end{equation}
To satisfy the second equation in Eq.(\ref{n-u-T}), we require
\begin{equation}
\sum\limits_{ki}{F_{k}v_{ki\alpha }}v_{ki\varepsilon }u_{\varepsilon
}=Tu_{\alpha }\text{,}  \label{FK-3}
\end{equation}%
\begin{equation}
\sum\limits_{ki}{F_{k}v_{ki\alpha }v_{ki\varepsilon}v_{ki\eta }v_{ki\xi
}u_{\varepsilon }u_{\eta }u_{\xi }}=3T^{2}u^{2}u_{\alpha}\text{.}
\label{FK-4}
\end{equation}
To satisfy the third equation in Eq.(\ref{n-u-T}), we require%
\begin{equation}
\sum\limits_{ki}{F_{k}\frac{v{_{k}^{2}}}{2}}=T\text{,}  \label{FK-5}
\end{equation}%
\begin{equation}
\sum\limits_{ki}{F_{k}\frac{{v_{k}^{2}}}{2}v_{ki\varepsilon }v_{ki\eta
}u_{\varepsilon }u_{\eta}}=2T^{2}u^{2}\text{.}  \label{FK-6}
\end{equation}
To satisfy Eq.(\ref{moment-4}), we require
\begin{equation}
\sum\limits_{ki}{F_{k}v_{ki\alpha }}v_{ki\beta}=T\delta_{\alpha \beta }\text{%
,}  \label{FK-7}
\end{equation}%
\begin{equation}
\sum\limits_{ki}{F_{k}v_{ki\alpha }v_{ki\beta }v_{ki\varepsilon }v_{ki\eta
}u_{\varepsilon }u_{\eta }}=T^{2}(u^{2}\delta_{\alpha \beta }+2u_{\alpha
}u_{\beta })\text{.}  \label{FK-8}
\end{equation}
To satisfy Eq.(\ref{moment-5}), we require
\begin{equation}
\sum\limits_{ki}{F_{k}\frac{{v_{k}^{2}}}{2}}v_{ki\alpha }v_{ki\varepsilon
}u_{\varepsilon }=2T^{2}u_{\alpha }\text{,}  \label{FK-9}
\end{equation}%
\begin{equation}
\sum\limits_{ki}{F_{k}\frac{{v_{k}^{2}}}{2}v_{ki\alpha }v_{ki\varepsilon
}v_{ki\eta }v_{ki\xi }u_{\varepsilon }u_{\eta }}u_{\xi
}=9T^{3}u^{2}u_{\alpha }\text{.}  \label{FK-10}
\end{equation}

If further consider the isotropic properties of the discrete
velocity model, the above ten requirements reduce to four ones.
Requirement Eq.(\ref{FK-1}) gives
\begin{equation}
\sum\limits_{ki}{F_{k}=1}\text{.}  \label{FK-11}
\end{equation}%
Requirements Eqs.(\ref{FK-2}),(\ref{FK-3}),(\ref{FK-5}),(\ref{FK-7}) give
\begin{equation}
\sum\limits_{k}{F_{k}v_{k}^{2}=}\frac{{T}}{3}\text{.}  \label{FK-12}
\end{equation}%
Requirements Eqs.(\ref{FK-4}),(\ref{FK-6}),(\ref{FK-8}),(\ref{FK-9}) give
\begin{equation}
\sum\limits_{k}{F_{k}v_{k}^{4}=}\frac{4}{3}{T}^{2}\text{.}  \label{FK-13}
\end{equation}%
Requirement Eq.(\ref{FK-10}) gives
\begin{equation}
\sum\limits_{k}{F_{k}v_{k}^{6}=}8{T}^{3}\text{.}  \label{FK-14}
\end{equation}

To satisfy the above four requirements, four particle velocities ${v_{k}}$
are necessary. We choose a zero speed ${v_{0}=0}$, and other three nonzero
ones. Eqs.(\ref{FK-11})-(\ref{FK-14}) are easily solved to the following
solution
\begin{equation}
F_{k}=\frac{{24T^{3}-4(v_{k+1}^{2}+v_{k+2}^{2})T^{2}+v_{k+1}^{2}v_{k+2}^{2}T}%
}{{3v_{k}^{2}(v_{k}^{2}-v_{k+1}^{2})(v_{k}^{2}-v_{k+2}^{2})}}\text{,}
\end{equation}%
\begin{equation}
F_{0}=1-6(F_{1}+F_{2}+F_{3})\text{,}
\end{equation}%
with
\begin{equation}
k+l=\left\{
\begin{array}{l}
k+l\text{, \ \ \ \ \textbf{\ }if }k+l\leq 3\text{;} \\
k+l-3\text{, if }k+l>3\text{.}%
\end{array}%
\right. l=1\text{, }2\text{.}
\end{equation}

The FDLB scheme breaks the binding of discretizations of space and
time and makes the particle speeds more flexible.
As far as a simulation being stably conducted, the specific values of ${v_{1}%
}$, ${v_{2}}$ and ${v_{3}}$ do not affect the accuracy of the
results. The flexibility can be used to obtain the temperature range
as wide as possible. In our simulations we set ${v_{1}=1}$,
${v_{2}=2}$ and ${v_{3}=3}$.

Moreover, it should be noted that the physical process described by
the LB model has an intrinsic viscosity which is linearly related to
the relaxation time $\tau$. The Euler equations or the Navier-Stokes
equations are only approximations of the LB modeling in the
continuum limit. In our case, we ensure only that the LB model
recovers the correct Euler equations when the Knudsen number is
zero.

\section{Spatiotemporal discretizations and verification of the model}

\subsection{Time and space discretizations}

In this section, we will confirm validity of the model by conducting
numerical simulations. Here the time derivative is calculated using
the forward Euler FD scheme. Spatial derivatives in convection term
$\mathbf{v}_{ki}.\frac{\partial }{{\partial \mathbf{r}}}f_{ki}$ are
calculated using fifth order Weighted Essentially Non-Oscillatory
(5th-WENO) FD scheme \cite{WENO-5th}.

The WENO scheme is an improvement of the Essentially Non-Oscillatory
(ENO) scheme, which changes the method of choosing smooth stencil
with logical judgment into weighted average of all stencils, thereby
improving the accuracy, the computational efficiency, and the
adaptability for compressible flows containing shock waves, contact
discontinuities, etc.

The basic idea of the ENO scheme is to use the ``smoothest" stencil
among several candidates to approximate the numerical fluxes to a
high order accuracy, and at the same time, to avoid spurious
oscillations near shocks. The ENO scheme is uniformly high order
accurate, robust, and there is essentially no oscillation near shock
waves \cite{ENO}. However, this scheme has several drawbacks that
can be improved by WENO scheme. For example, compared to WENO
scheme, the ENO scheme is less accurate in smooth regions, and the
numerical flux is not as smooth as that of the WENO scheme. For WENO
scheme, instead of approximating the numerical flux using only one
of the candidate stencils, it uses all candidate stencils through a
convex combination. The contribution of each stencil to the final
approximation of the numerical flux is determined by a weight, which
can be defined in such a way that in smooth regions it approaches
certain optimal weights to achieve a higher order of accuracy, while
in regions near discontinuities, the stencils which contain the
discontinuities are assigned a nearly zero weight \cite{WENO-5th}.
As a result, the accuracy can be improved to the optimal order (in
Ref. \cite{WENO-4th}, a $r$th order ENO scheme has been improved to
a ($r+1$)th WENO scheme, and in Ref. \cite{WENO-5th} the ($r+1$)th
WENO scheme has been further improved to a ($2r-1$)th WENO scheme)
in smooth regions and the essentially non-oscillatory property near
discontinuities can be maintained.

Now we spell out the details of the 5th-WENO scheme. To be specific,
the discrete evolution equation of distribution function
$f_{ki,I}^{n+1}$ is
\begin{eqnarray}
f_{ki,I}^{n+1} &=&f_{ki,I}^{n}-\frac{{\partial (v_{ki\alpha }f_{ki,I}^{n})}}{%
{\partial r_{\alpha }}}\Delta t-\frac{{1}}{\tau }(f_{ki,I}^{n}-f_{ki,I}^{n,{%
(0)}})\Delta t  \notag \\
\text{\ } &{=}&f_{ki,I}^{n}-\frac{1}{{\Delta r_{\alpha }}}\left( {%
h_{i,I+1/2}-h_{i,I-1/2}}\right) \Delta t-\frac{{1}}{\tau }%
(f_{ki,I}^{n}-f_{ki,I}^{n,{(0)}})\Delta t\text{,}
\end{eqnarray}%
where the superscripts $n,n+1$ indicate the consecutive two
iteration steps,
$I$ is the index of lattice node, $\Delta t$ is the time step, and ${%
h_{i,I+1/2}}$ is the numerical flux and can be defined as a convex
combination
\begin{equation}
h_{i,I+1/2}=\omega _{1}h_{i,I+1/2}^{1}+\omega _{2}h_{i,I+1/2}^{2}+\omega
_{3}h_{i,I+1/2}^{3}\text{.}  \label{Flux}
\end{equation}

Under the condition ${v_{ki\alpha }\geq 0}$, interpolation functions $%
h_{i,I+1/2}^{q}(q=1$, $2$, $3)$ are given by
\begin{equation}
h_{i,I+1/2}^{1}=\frac{1}{3}F_{i,I-2}-\frac{7}{6}F_{i,I-1}+\frac{{11}}{6}%
F_{i,I}\text{,}
\end{equation}%
\begin{equation}
h_{i,I+1/2}^{2}=-\frac{1}{6}F_{i,I-1}+\frac{5}{6}F_{i,I}+\frac{1}{3}F_{i,I+1}%
\text{,}
\end{equation}%
\begin{equation}
h_{i,I+1/2}^{3}=\frac{1}{3}F_{i,I}+\frac{5}{6}F_{i,I+1}-\frac{1}{6}F_{i,I+2}%
\text{,}
\end{equation}%
where $F_{i,I}={v_{ki\alpha }f_{ki,I}}$.

The weighting factors $\omega _{q}$ reflect the smooth degree of the
template. They can be defined as follows
\begin{equation}
\omega _{q}=\frac{{\tilde{\omega}_{q}}}{{\tilde{\omega}_{1}+\tilde{\omega}%
_{2}+\tilde{\omega}_{3}}}\text{, }\tilde{\omega}_{q}=\frac{{\delta _{q}}}{{%
\left( {10^{-6}+\sigma_{q}}\right) ^{2}}}\text{,}  \label{wq}
\end{equation}%
where ${\delta _{1}=1/10}$, ${\delta _{2}=3/5}$ and ${\delta
_{3}=3/10}$ are the optimal weights. The small value ${10^{-6}}$ is
added to the denominator to avoid dividing by zero. The coefficients
${\sigma _{q}}$ in Eq.(\ref{wq}) are the smoothness
indicators, and can be obtained by%
\begin{equation}
\sigma _{1}=\frac{{13}}{{12}}(F_{i,I-2}-2F_{i,I-1}+F_{i,I})^{2}+\frac{1}{4}%
(F_{i,I-2}-4F_{i,I-1}+3F_{i,I})^{2}\text{,}
\end{equation}%
\begin{equation}
\sigma _{2}=\frac{{13}}{{12}}(F_{i,I-1}-2F_{i,I}+F_{i,I+1})^{2}+\frac{1}{4}%
(F_{i,I-1}-F_{i,I+1})^{2}\text{,}
\end{equation}%
\begin{equation}
\sigma _{3}=\frac{{13}}{{12}}(F_{i,I}-2F_{i,I+1}+F_{i,I+2})^{2}+\frac{1}{4}%
(3F_{i,I}-4F_{i,I+1}+F_{i,I+2})^{2}\text{.}  \label{D-3}
\end{equation}

Similarly, under the condition ${v_{ki\alpha }<0}$, a mirror image procedure
of the procedure from Eqs.(\ref{Flux}) to (\ref{D-3}) can be carried out.

\subsection{Numerical test: one-dimensional Sod shock tube}

The validity of the formulated LB model is verified through a
one-dimensional Riemann problem, the Sod shock tube \cite{SOD}. This is a
classical test in the research of compressible flows, which contains the
shock wave, the rarefaction wave and the contact discontinuity.

%%%%%%%%%%%%%%%%%%%%%%%%%%%%%%%%%%%%%%%%%%%%%%%%%%%%%%%%%%%%%%%%%%%%%%
\begin{figure}[tbp]
\center{\epsfig{file=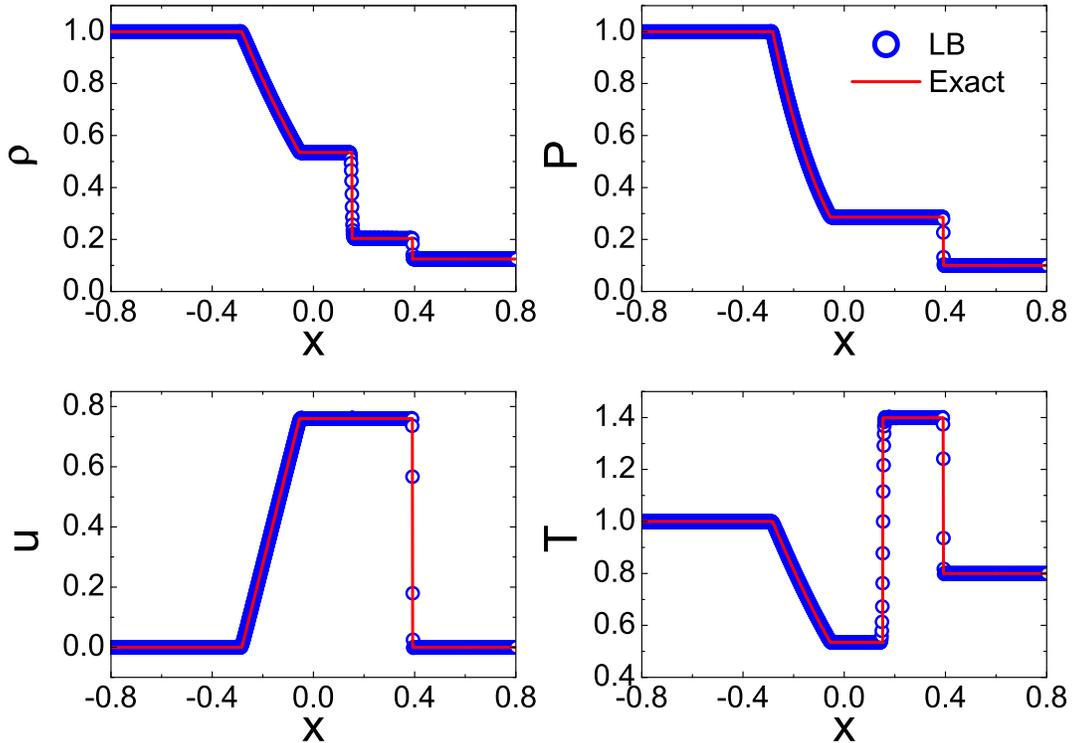,bbllx=0pt,bblly=0pt,bburx=434pt,bbury=328pt,
width=0.88\textwidth,clip=}} \caption{(Color online) Comparisons
between LB results and exact ones for the one-dimensional Sod
problem, where $t=0.2$.}
\end{figure}
%%%%%%%%%%%%%%%%%%%%%%%%%%%%%%%%%%%%%%%%%%%%%%%%%%%%%%%%%%%%%%%%%%%%%

For the problem considered, the initial condition is described by
\begin{equation}
\left\{
\begin{array}{l}
\left. {(\rho }\text{{, }}{u}\text{{, }}{v}\text{{, }}{T)}\right\vert
_{L}=(1.0\text{, }0.0\text{, }0.0\text{, }1.0)\text{{,} \ \ \ }x\leq 0\text{;%
} \\
\left. {(\rho }\text{{, }}{u}\text{{, }}{v}\text{{, }}{T)}\right\vert
_{R}=(0.125\text{, }0.0\text{, }0.0\text{, }0.8)\text{, \ }x>0\text{.}%
\end{array}%
\right.
\end{equation}%
Subscript  ``$L$" and ``$R$" indicate macroscopic variables at the
left and right sides of the discontinuity. The size of grid is
$\Delta x=\Delta y=10^{-3}$, time step and relaxation time are
$\Delta t=\tau =10^{-5}$.

The periodic boundary conditions are taken
in the $y$ direction. In the $x$ direction, we set
\begin{equation}
f_{ki\text{,}-2\text{,}t}=f_{ki\text{,}-1\text{,}t}=f_{ki\text{,}0\text{,}%
t}=f_{ki\text{,}1\text{,}t=0}^{(0)}\text{,}  \label{mic-left}
\end{equation}%
where $-2$, $-1$ and $0$ are the indexes of ghost nodes on the left
side. Such a boundary condition means that the system at the
boundaries keeps at their corresponding equilibrium states and the
macroscopic quantities on the boundary nodes keep their initial
values \cite{QuKun,HYL},
\begin{equation}
(\rho \text{, }u\text{, }v\text{, }P)_{I=-2\text{,}t}=(\rho \text{, }u\text{%
, }v\text{, }P)_{I=-1\text{,}t}=(\rho \text{, }u\text{, }v\text{, }P)_{I=0%
\text{,}t}=(\rho \text{, }u\text{, }v\text{,
}P)_{I=1\text{,}t=0}\text{.} \label{mac-left}
\end{equation}%
On the right boundary, we can do in a similar way. The boundary
conditions in the $x$-direction are also referred to as the
supersonic inflow boundary conditions \cite{QuKun}.
Eq.(\ref{mic-left}) and Eq.(\ref{mac-left}) may be referred to as
the microscopic and the macroscopic boundary conditions,
respectively, which are consistent with each other.

It is worth noting that when the external environment is in
equilibrium the boundary conditions for the FDLB model are
essentially the same as those in traditional computational fluid
dynamics (CFD). When the external
environment is not in equilibrium, the non-equilibrium part $f_{ki\text{,}%
I}^{(neq)}$ can be obtained from the inner lattice nodes via the
extrapolation method. That nonequilibrium effects can be
incorporated from boundary conditions is a merit of LB method over
the traditional CFD.

%%%%%%%%%%%%%%%%%%%%%%%%%%%%%%%%%%%%%%%%%%%%%%%%%%%%%%%%%%%%%%%%%%%%%%
\begin{figure}[tbp]
\center{\epsfig{file=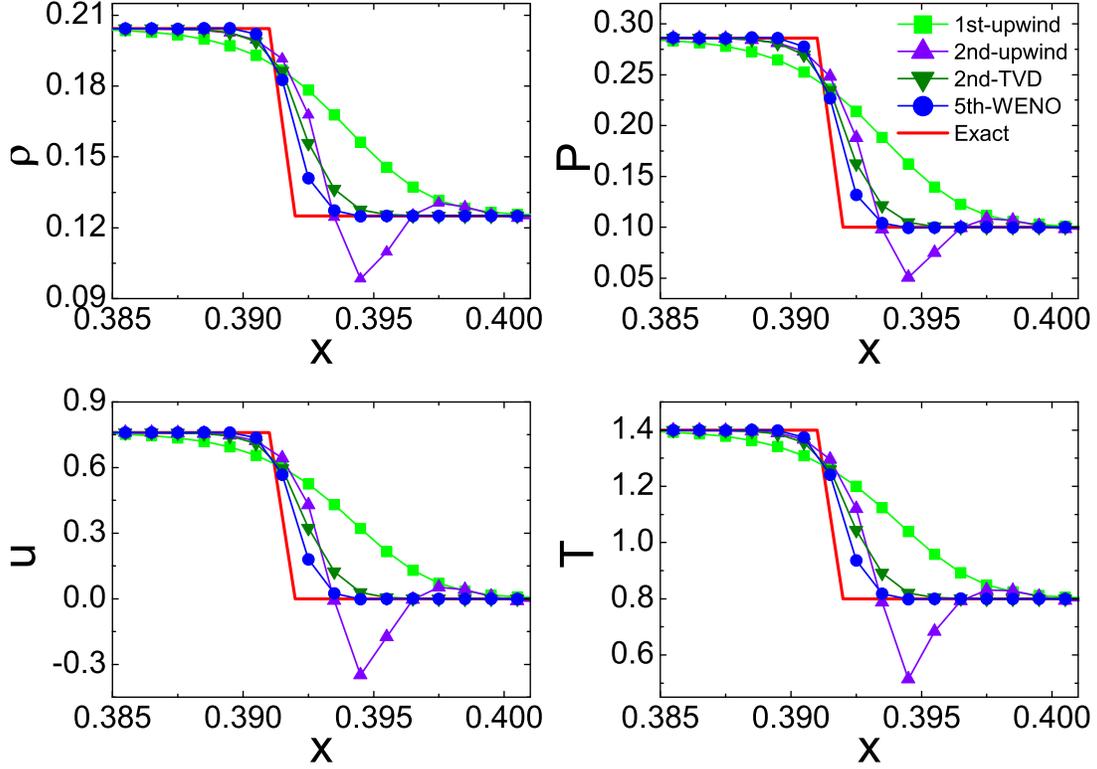,bbllx=8pt,bblly=3pt,bburx=535pt,bbury=390pt,
width=0.88\textwidth,clip=}} \caption{(Color online) Local details
of the profiles near the shock wave for Sod shock problem at
$t=0.2$. The 1st-upwind scheme, the 2nd-upwind scheme, the 2nd-TVD
scheme, and the 5th-WENO scheme are adopted in spatial
discretization for comparisons. }
\end{figure}
%%%%%%%%%%%%%%%%%%%%%%%%%%%%%%%%%%%%%%%%%%%%%%%%%%%%%%%%%%%%%%%%%%%%%
Figure 1 shows the computed density, pressure, velocity, and
temperature profiles at $t=0.2$, where the circles are for
simulation results and solid lines are for analytical solutions. The
two sets of results have a satisfying agreement. Figure 2 enlarges
the part containing the shock wave for a closing view. The
first-order upwind (1st-upwind) scheme, the second order upwind
(2nd-upwind) scheme, the second order total variation diminishing
(2nd-TVD) scheme \cite{NND}, and the 5th-WENO scheme are adopted in
spatial discretization for comparisons. One can see that the
1st-upwind scheme has a strong ``smoothing effect" and gives a very
smeared solution with excessive numerical dissipation. Compared to
the former, the 2nd-upwind scheme has a weak ``smoothing effect",
but it introduces the unphysical oscillations at discontinuities.
The 2nd-TVD scheme effectively eliminates the spurious numerical
oscillations at discontinuities, while the numerical dissipation may
need to be reduced further. However, with the 5th-WENO scheme, not
only the spurious numerical oscillations at discontinuities are
effectively refrained but also the numerical dissipation is severely
curtailed. With an increase in the order of the schemes, fewer nodes
are needed to capture the shock waves. The widths of the shock waves
are spread over only about three to four grid cells with the
5th-WENO scheme, which shows that the LB model with the WENO scheme
has a high ability to capture shocks for this test.

\section{Effects of velocity and density gradients on KHI}

At the initial linear increasing stage of KHI, the amplitude $\eta$
of perturbation evolves according to the following relation, $\eta
=\eta _{0}e^{\gamma t}$, where $\gamma$ is the growth coefficient
and is dependent on the gradient of tangential velocity and gradient
of density around the interface \cite{ICF-book}. In other words,
$\gamma$ is dependent on the width of velocity transition layer $D_v
$ and width of density transition layer $D_{\rho}$. In this section,
we discuss separately the KHI in three cases, (i) $D_v$ is variable
and $D_{\rho}$ is fixed, (ii) $D_{\rho}$ is variable and $D_v$ is
fixed, (iii) both $D_{\rho}$ and $D_{v}$ are variable. The
increasing rate $\gamma$ for case (i), (ii) and (iii) are referred
to as $\gamma_v$, $\gamma_{\rho}$ and $\gamma_R$, respectively. We
numerically obtain $\gamma_v$, $\gamma_{\rho}$
 and $\gamma_R$ via fitting the curves of $%
\ln E_x |_{\max} (t)$ versus the time $t$, where $E_x |_{\max} (t)$
is the maximum of $E_x (x, y, t)$ in the whole computational domain,
$E_{x} (x,y,t) = \rho (x,y,t) u^2(x,y,t)/2$ is the perturbed kinetic
energy at the position ($x$, $y$) at each time step $t$.

\subsection{Linear growth rate and velocity gradient effect}

The initial configurations in our simulation are described by
\begin{equation}
\rho (x)=\frac{{\rho _{L}+\rho _{R}}}{2}-\frac{{\rho _{L}-\rho _{R}}}{2}%
\tanh (\frac{x}{{D_{\rho }}})\text{,}
\end{equation}%
\begin{equation}
v(x)=\frac{{v_{L}+v_{R}}}{2}-\frac{{v_{L}-v_{R}}}{2}\tanh (\frac{x}{{D_{v}}})%
\text{,}
\end{equation}%
\begin{equation}
P_{L}=P_{R}=P\text{,}
\end{equation}%
where ${D_{\rho }}$ and ${D_{v}}$ are the widths of density and velocity
transition layers. The velocity(density) is discontinuous at $x=0$ when ${%
D_{v}=0}$(${D_{\rho }=0}$). ${\rho _{L}=5.0}$(${\rho _{R}=2.0}$) is the
density away from the interface of the left(right) fluid. ${v_{L}=0.5}$(${%
v_{R}=-0.5}$) is the velocity away from the interface in
$y$-direction of the left(right) fluid and $P_{L}$($P_{R}$)$=2.5$ is
the pressure in the left(right) side. The Mach numbers of the left
side and the right side of the simulation domain are  $0.5$ and
$0.32$, respectively. Hence the flow speed $v_0$ is subsonic and the
system  can be approximately thought of as ``incompressible".
The whole calculation domain is a rectangle with length $%
0.6$ and height $0.2$, which is divided into $600\times 200$ uniform
meshes. A simple velocity perturbation in the $x$-direction is
introduced to trigger the KH rollup\textbf{\ }and it is in the
following form
\begin{equation}
u=u_{0}\sin (ky)\exp (-kx)\text{,}  \label{uu}
\end{equation}%
where $u_{0}=0.02$ is the amplitude of the perturbation. Here, $k$
is the wave number of the initial perturbation, and is set to be
$10\pi $. The time step is $\Delta t=10^{-5}$. Extensive studies
indicate that viscosity will damp the evolution of the KHI. However,
for the KHI in some cases, for example, in ICF, effects of the
viscosity are generally negligible. Therefore, throughout the
simulations, $\tau$ is set to be $10^{-5}$ to reduce the physical
viscosity. Boundary conditions for the simulations of KHI are as
below. Periodic boundary conditions are used in the $y$-direction.
In the horizontal direction, we adopt the outflow (zero gradient)
boundary conditions, which are widely used by other authors to
simulate the KHI \cite{outflow-KHI-1,outflow-KHI-2,outflow-KHI-3}.
For the outflow boundary conditions, the zeroth-order extrapolation
is used \cite{QuKun}. According to this boundary conditions, at the
left side, we set
\begin{equation}
(\rho \text{, }u\text{, }v\text{, }P)_{I=-2\text{,}t}=(\rho \text{, }u\text{%
, }v\text{, }P)_{I=-1\text{,}t}=(\rho \text{, }u\text{, }v\text{, }P)_{I=0%
\text{,}t}=(\rho \text{, }u\text{, }v\text{, }P)_{I=1\text{,}t}\text{.}
\label{mac-left-2}
\end{equation}%
and at the right side
\begin{equation}
(\rho \text{, }u\text{, }v\text{, }P)_{I=N_{x}+3\text{,}t}=(\rho \text{, }u%
\text{, }v\text{, }P)_{I=N_{x}+2\text{,}t}=(\rho \text{, }u\text{, }v\text{,
}P)_{I=N_{x}+1\text{,}t}=(\rho \text{, }u\text{, }v\text{, }P)_{I=N_{x}\text{%
,}t}\text{,}  \label{mac-right-2}
\end{equation}%
where $I=-2$, $-1$, $0$ and $N_{x}+1$, $N_{x}+2$, $N_{x}+3$ are the
indexes of left and the right ghost nodes, respectively.

%%%%%%%%%%%%%%%%%%%%%%%%%%%%%%%%%%%%%%%%%%%%%%%%%%%%%%%%%%%%%%%%%%%%%%%%%
\begin{figure}[tbp]
\center{\epsfig{file=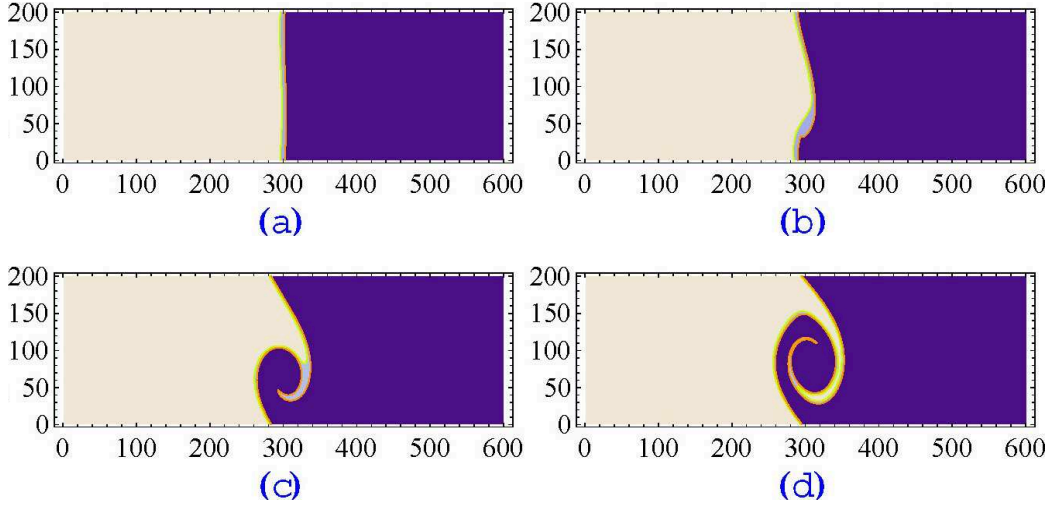,bbllx=123pt,bblly=588pt,bburx=498pt,bbury=775pt,
width=0.85\textwidth,clip=}} \caption{(Color online) Density
evolutions of KHI simulated using the LB model, where $D_v = 4$ and
${D_{\protect\rho }=8}$, $t = 0.1$ in (a), $t = 0.3$ in (b), $t =
0.5$ in (c), and $t = 0.7$ in (d). }
\end{figure}
%%%%%%%%%%%%%%%%%%%%%%%%%%%%%%%%%%%%%%%%%%%%%%%%%%%%%%%%%%%%%%%%%%%%%%%%%%%%

Figure 3 shows the temporal evolution of the density field for
${D_{v}=4}$ and ${D_{\rho }=8}$ at four different times. It is clear
that at $t=0.3$ the interface is wiggling due to the initial
perturbation and the velocity shear. After the initial linear growth
stage, there is a nicely rolled vortex developing around the initial
interface. A larger vortex is observed in the snapshot at $t=0.7$,
and a mixing layer is expected to be formed around the initial
interface. The interface is continuous and smooth, which indicates
the LB model has a good capturing ability of interface deformation.

%%%%%%%%%%%%%%%%%%%%%%%%%%%%%%%%%%%%%%%%%%%%%%%%%%%%%%%%%%%%%%%%%%%%%%
\begin{figure}[tbp]
\center{\epsfig{file=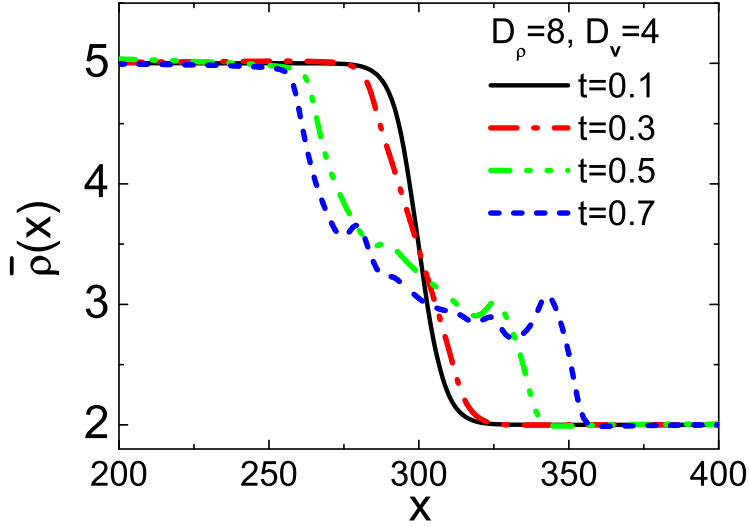,bbllx=0pt,bblly=0pt,bburx=432pt,bbury=321pt,
width=0.6\textwidth,clip=}}
\caption{(Color online) Averaged density profiles along the $x$-axis at $%
t=0.1$ (solid line), $t=0.3$ (dash dot line), $t=0.5$ (dash dot dot line),
and $t=0.7$ (short dash line) for ${D_{\protect\rho}=8}$ and ${D_{v}=4}$.}
\end{figure}
%%%%%%%%%%%%%%%%%%%%%%%%%%%%%%%%%%%%%%%%%%%%%%%%%%%%%%%%%%%%%%%%%%%%%
To quantitatively describe the characteristics of the vortex or the
mixing layer, in Fig.4 we plot the averaged density profile
$\overline{\rho }(x)$ against the $x$-axis at $t=0.1$, $0.3$, $0.5$
and $0.7$. The averaged density in the mixing layer is defined as
\begin{equation}
\overline{\rho }(x)=\frac{1}{L}\int_{0}^{L}{\rho (x,y)dy}\text{.}
\end{equation}%
The averaged density profiles vary from being smooth to being irregular. The
thickness of the mixing layer and the amplitude of the density oscillation
increase with time. The zig-zags in the profiles indicate the transfer of
fluids from the dense to the rarefactive regions and the irregularity in the
mixing layer.
%%%%%%%%%%%%%%%%%%%%%%%%%%%%%%%%%%%%%%%%%%%%%%%%%%%%%%%%%%%%%%%%%%%%%%%%%%
\begin{figure}[tbp]
\center{\epsfig{file=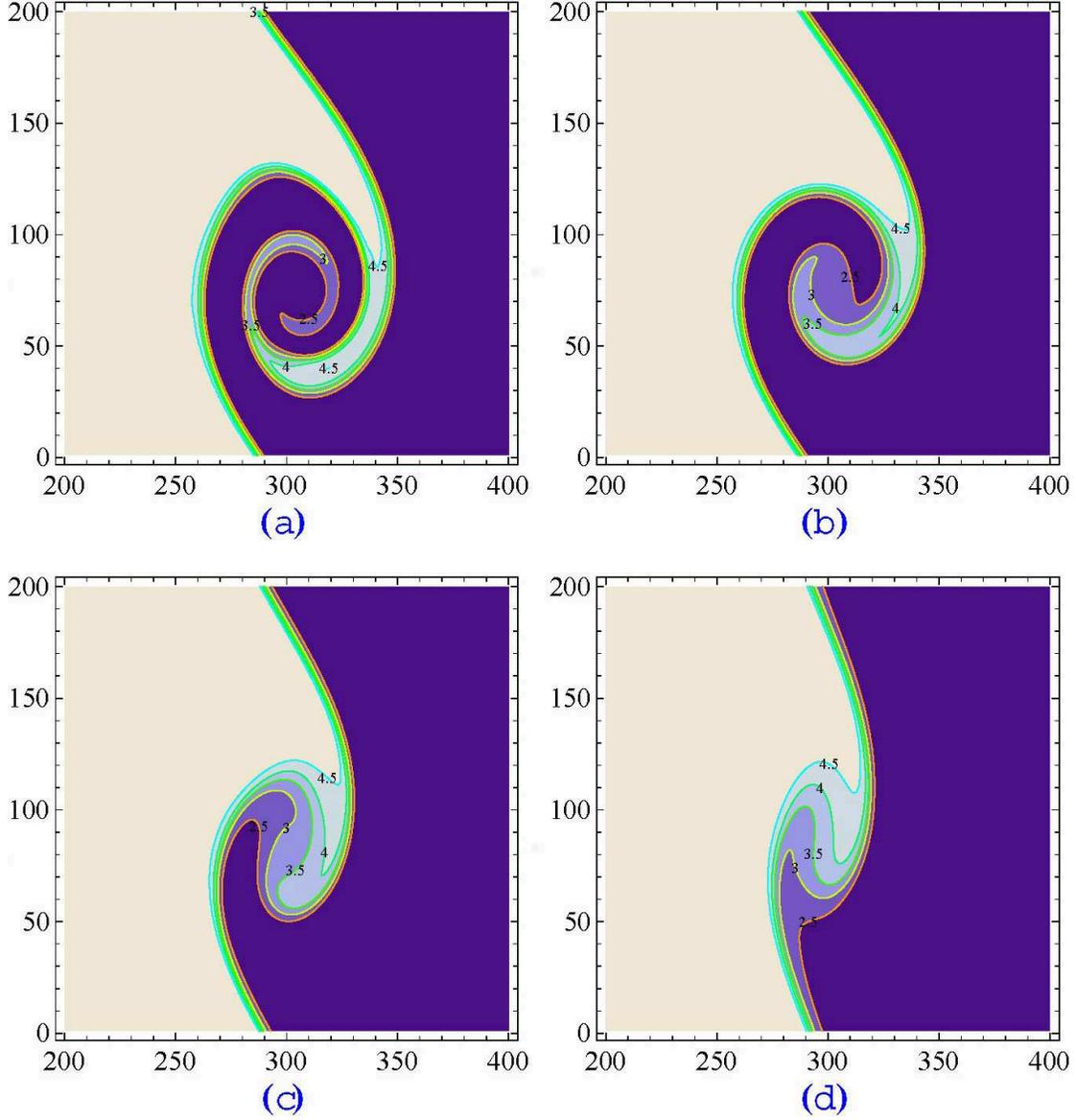,bbllx=72pt,bblly=264pt,bburx=551pt,bbury=770pt,
width=0.9\textwidth,clip=}}
\caption{(Color online) Vortices in the mixing layer as a function of ${D_{v}%
}$ at $t=0.6$, where ${D_{v}=4}$ in (a), ${D_{v}=8}$ in (b), ${D_{v}=12}$ in
(c), and ${D_{v}=16}$ in (d). The density transition layer ${D_{\protect\rho %
}}$ is fixed to be $8$. }
\end{figure}
%%%%%%%%%%%%%%%%%%%%%%%%%%%%%%%%%%%%%%%%%%%%%%%%%%%%%%%%%%%%%%%%%%%%%%%%%%%%
%%%%%%%%%%%%%%%%%%%%%%%%%%%%%%%%%%%%%%%%%%%%%%%%%%%%%%%%%%%%%%%%%%%%%%
\begin{figure}[tbp]
\center{\epsfig{file=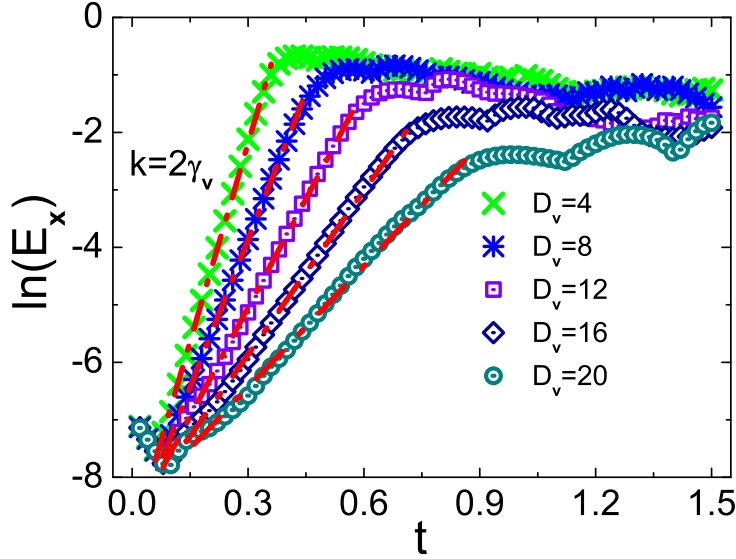,bbllx=0pt,bblly=0pt,bburx=436pt,bbury=332pt,
width=0.6\textwidth,clip=}} \caption{(Color online) Time evolution
of the perturbed peak kinetic energy $E_x |_{\max}$ along the
$x$-axis in $\ln $-linear scale for various widths of velocity
transition layer. The dash-dotted lines represent the linear fits to
the initial linear growth regimes.}
\end{figure}
%%%%%%%%%%%%%%%%%%%%%%%%%%%%%%%%%%%%%%%%%%%%%%%%%%%%%%%%%%%%%%%%%%%%%%
\begin{figure}[tbp]
\center{\epsfig{file=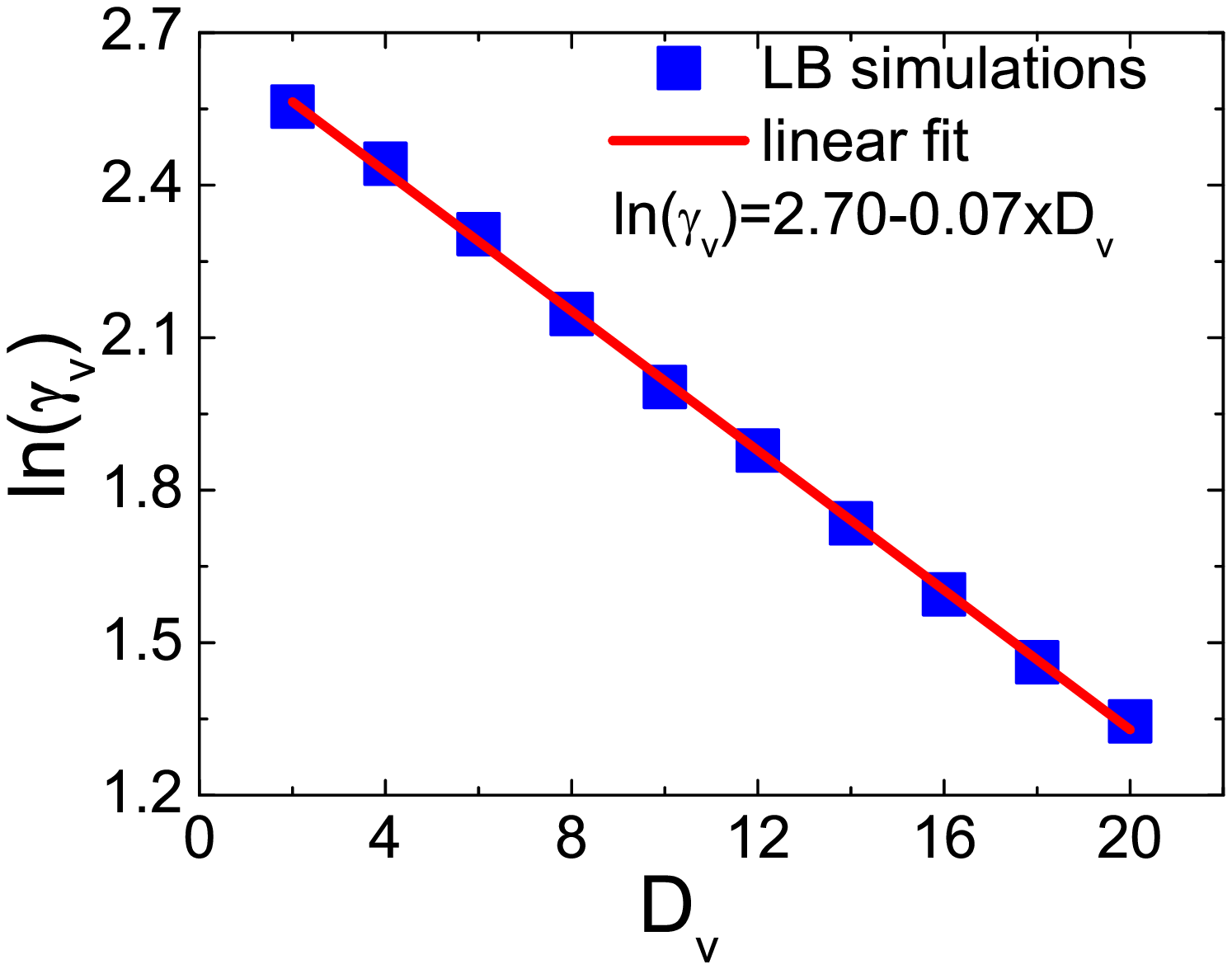,bbllx=0pt,bblly=0pt,bburx=444pt,bbury=342pt,
width=0.6\textwidth,clip=}}
\caption{(Color online) Linear growth rate as a function of the width ${D_{v}%
}$ of velocity transition layer. }
\end{figure}
%%%%%%%%%%%%%%%%%%%%%%%%%%%%%%%%%%%%%%%%%%%%%%%%%%%%%%%%%%%%%%%%%%%%%

To consider the velocity gradient effect, the simulations with various ${%
D_{v}}$ have been performed and the density maps for ${D_{v}=4}${, }${8}${, }%
${12}$, and $16$ with five contour lines at $t=0.6$ are plotted in
Fig.5. We see that the width of the velocity transition layer can
significantly affect the evolution of KHI. For fixed density
gradient and velocity difference, the larger the value of $D_{v}$,
i.e., the wider the initial transition zone, the weaker the KHI, and
the later the vortex appears. In Fig.5(a) and Fig.5(b), the
appearance of large vortices demonstrates that the evolution is
embarking on the nonlinear stage. While in Fig.5(c) and Fig.5(d),
the width of the mixing layers is rapidly decreased by the increase
of ${D_{v}}$ and the evolution is in the weakly nonlinear stage.
Figures 5(a)-(d) show that a wider velocity transition zone has a
better stabilization effect on KHI.

Taking the logarithm of the perturbed peak kinetic energy $E_x
|_{\max}$ of each time step, the exponential growth manifests itself
as a linear slope \cite{ek,ek2,ek3} (see Fig.6). Therefore, $\gamma$
can be obtained from the slope of a linear function fitted to the
initial growth stage. We note in this respect that $E_{x}\propto $
$u^{2}\propto \left(
e^{\gamma t}\right) ^{2}$ grows at twice the rate of the KHI, namely $%
k=2\gamma $, where $k$ is the slope of the linear fit. The peak
kinetic energy $E_x |_{\max}$ can represent the interacting strength
of two different fluids. It is clear from Fig.6 that the logarithm
of the perturbed kinetic energy first grows linearly in time and
then arrives at a nonlinear saturation procedure at late times.

For a fixed ${D_{v}}$ and ${D_{\rho}}$, $E_x |_{\max}$ increases
with time $t$ during the linear growth stage. However, at the same
time, the larger the value of $D_{v}$, the smaller the perturbed
peak kinetic energy $E_x |_{\max}$. This means when the transition
layer becomes wider, the evolution of the
density field becomes slower. Moreover, we find that the dependence of $%
\gamma _{v}$ on $D_{v}$ can be fitted by a logarithmic function with the
form
\begin{equation}
\ln \gamma _{v}=a-bD_{v}\text{,}  \label{vv}
\end{equation}%
with $a=2.70$ and $b=0.07$ as displayed in Fig.7. The numerical
simulation result is in general agreement with the analytical
results (see Eq.(18) and Fig.3 in recent work of Wang, \emph{et al.}
\cite{wang-pop-2010}). In the classical
case, the linear growth rate is $\gamma _{c}=k\sqrt{\rho _{1}\rho _{2}}%
(v_{1}-v_{2})/(\rho _{1}+\rho _{2})\propto \Delta v$, where $\Delta
v$ is the shear velocity difference. A wider transition layer
decreases the local or the effective shear velocity difference
$\Delta {v}$, which results in a smaller linear growth rate, a
smaller saturation energy, and a longer linear growth time
$t_{lin}$.

\subsection{Linear growth rate and density gradient effect}

%%%%%%%%%%%%%%%%%%%%%%%%%%%%%%%%%%%%%%%%%%%%%%%%%%%%%%%%%%%%%%%%%%%%%%
\begin{figure}[tbp]
\center{\epsfig{file=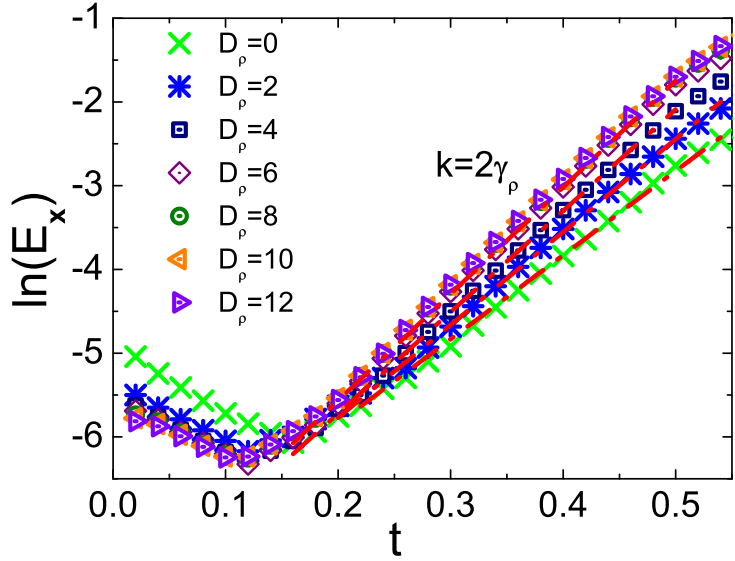,bbllx=0pt,bblly=0pt,bburx=439pt,bbury=334pt,
width=0.6\textwidth,clip=}} \caption{(Color online) Time evolution
of the logarithm of the peak kinetic energy $E_x |_{\max}$ along the
$x$-axis for various widths of density transition layers.}
\end{figure}
%%%%%%%%%%%%%%%%%%%%%%%%%%%%%%%%%%%%%%%%%%%%%%%%%%%%%%%%%%%%%%%%%%%%%
%%%%%%%%%%%%%%%%%%%%%%%%%%%%%%%%%%%%%%%%%%%%%%%%%%%%%%%%%%%%%%%%%%%%%%
\begin{figure}[tbp]
\center{\epsfig{file=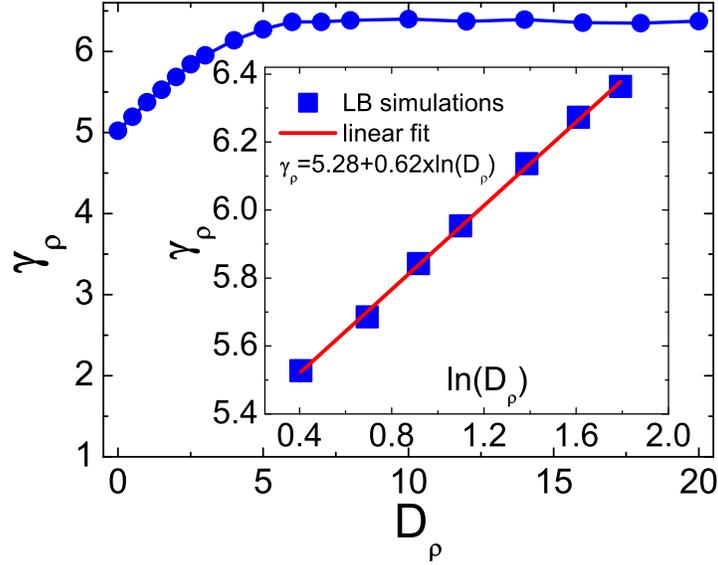,bbllx=0pt,bblly=0pt,bburx=581pt,bbury=445pt,
width=0.6\textwidth,clip=}}
\caption{(Color online) Linear growth rate as a function of the width ${D}_{%
\protect\rho }$ of density transition layer. }
\end{figure}
%%%%%%%%%%%%%%%%%%%%%%%%%%%%%%%%%%%%%%%%%%%%%%%%%%%%%%%%%%%%%%%%%%%%%

In this subsection, the density gradient effect is investigated in a
similar way. The initial conditions are described as, $({\rho_{L}}${, }${v_{L}}${, }$%
P_{L})=(5.0$, $0.5$, $1.5)$ and $({\rho_{R}}${, }${v_{R}}${,} $P_{R})=(1.25 $%
, $-0.5$, $1.5)$. Parameters are set to be $dx=dy=0.002$, $\Delta
t=10^{-5}$. Figure 8 shows time evolution of the logarithm of the
peak kinetic energy $E_x |_{\max}$ along the $x$-axis with various
widths of density transition layers. It is seen in Fig.8, for fixed
width $D_{v}$ of velocity transition layer and fixed density
difference, the linear growth rate first increases with the width
$D_{\rho}$ of density transition layer. But when ${D_{\rho}}>6$, the
linear growth rate does not vary significantly any more (see Fig.8
for details), which indicates the effective interaction width of
$D_{\rho}$ is less than that of $D_{v}$. Moreover, we find when
${D_{\rho}}<6$, the dependence of $\gamma_{\rho}$ on $D_{\rho}$ can
be fitted by the following equation
\begin{equation}
\gamma_{\rho}=c+e\ln D_{\rho}\text{,}  \label{rhorho}
\end{equation}%
with $c=5.28$ and $e=0.62$ as shown in the legend of Fig.9. The
numerical simulation results are also in general agreement with the
analytical results (see Fig.2 in previous work by Wang, \emph{et
al.} \cite{wang-pop-2009}, Eq.(18) and Fig.2 in recent work of Wang,
\emph{et al.} \cite{wang-pop-2010}). In the classical case, the
square of the linear growth rate is $\gamma_{c}^{2}=k^{2}\rho
_{1}\rho _{2}(v_{1}-v_{2})^{2}/(\rho_{1}+\rho_{2})^{2}\propto
(1-A^{2})\Delta v^{2}$, where $A=(\rho_{1}-\rho_{2})/(\rho_{1}+\rho
_{2})$ is the Atwood number. A wider density transition zone reduces
the Atwood number around the interface. Then in the process of
exchanging momentum in the direction normal to the interface, the
perturbation can obtain more energy from the shear kinetic energy
than in cases with sharper interfaces. Therefore, a wider density
transition layer increases the linear growth rate of the KHI.

\subsection{Hybrid effects of velocity and density gradients}

In practical systems, at the interface of two fluids with a tangential
velocity difference, both the velocity and the density gradients exist. The
wider velocity transition layer decreases the linear growth rate of the KHI,
but the wider density transition layer strengthens it. Consequently, there
is a competition between these two gradient effects. In this section, we
investigate how these two effects compete with each other. For convenience,
we introduce a coefficient $R=D_{\rho}/D_{v}$ to analyze the combined
effect. The linear growth rate at various velocity and density gradients for
$R=0.5$, $1$, $2$, and $5$ are illustrated in Fig.10.
%%%%%%%%%%%%%%%%%%%%%%%%%%%%%%%%%%%%%%%%%%%%%%%%%%%%%%%%%%%%%%%%%%%%%%
\begin{figure}[tbp]
\center{\epsfig{file=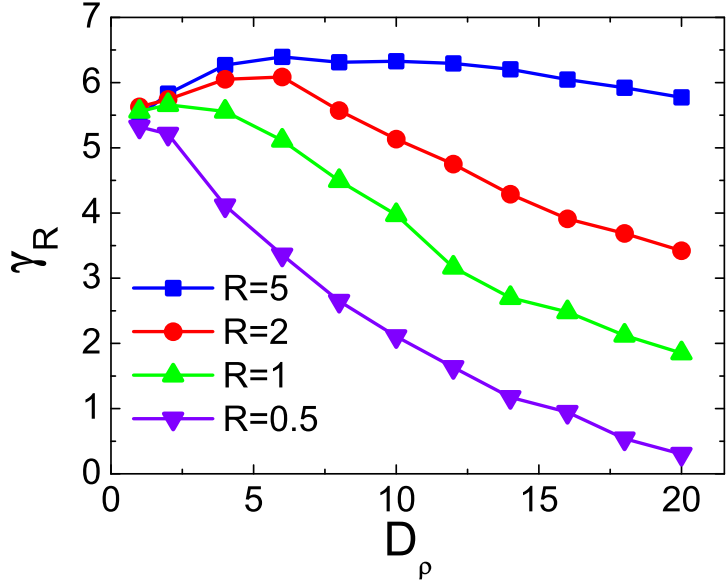,bbllx=14pt,bblly=0pt,bburx=581pt,bbury=451pt,
width=0.6\textwidth,clip=}} \caption{(Color online) The linear
growth rate versus the width of density transition layer for
$R=0.5$, $1$, $2$, and $5$. The initial density, shear
velocity and pressure of the two fluids are $({\protect\rho _{L}}${, }${v_{L}%
}${, }$P_{L})=(5.0$, $0.5$, $1.5)$ and $({\protect\rho _{R}}${, }${v_{R}}${,
}$P_{R})=(1.25$, $-0.5$, $1.5).$ }
\end{figure}
%%%%%%%%%%%%%%%%%%%%%%%%%%%%%%%%%%%%%%%%%%%%%%%%%%%%%%%%%%%%%%%%%%%%%

As shown in Fig.10, on the whole, the hybrid effect of velocity and
density transition layers reduces the linear growth rate $\gamma_R$.
Only at small $D_{\rho }$ and when $R>1$, the hybrid effect of
velocity and density transition layers makes larger the linear
growth rate. This indicates that the effective interaction width of
the velocity transition layer ${D_{v}^{E}}$ is wider than that of
density transition layer ${D_{\rho}^{E}}$.

For a fixed value of $R$ ($>1$), when $D_{\rho }$ is small, the transfer
efficiency of kinetic energy between neighboring layers increases with
decreasing the density gradient, so the linear growth rate increases with $%
D_{\rho }$ and reaches a peak. When $D_{\rho }$ is greater than the
peak value, the stabilizing effect of velocity transition layer
becomes more significant and stronger than the destabilizing effect
of the density transition layer, which leads to the decreasing of
linear growth rate with $D_{\rho }$, as shown in Fig.10 of $R=1$,
$2$, and $5$ cases. However, when $R<1$, the linear growth rate
decreases monotonically with $D_{\rho }$, no matter it is small or
large.

\section{Conclusions and discussions}

In this paper, a two-dimensional LB model with 19 discrete
velocities in six directions is proposed. The model allows to
recover the compressible Euler equations in the continuum limit.
With the introduction of the 5th-WENO FD scheme, the unphysical
oscillations at discontinuities are effectively suppressed and the
numerical dissipation is severely curtailed. The validity of the
model is verified by its application to the Sod shock tube, and
excellent agreement between the simulation results and the
analytical solutions can be found.

Using the proposed LB model, the velocity and density gradient
effects on the KHI have been studied. The averaged density profiles,
and the peak kinetic energy are used to quantitatively describe the
evolution of KHI. It is found that evolution of the KHI is damped
with increasing the width of velocity transition layer but is
strengthened with increasing the width of density transition layer.
After the initial transient period and before the vortex
has been well formed, the linear growth rates, $\gamma_v$ and $\gamma_{\rho}$%
, vary with ${D_{v}}$ and ${D_{\rho}}$ approximately in the following way, $%
\ln\gamma _{v}=a-bD_{v}$ and $\gamma _{\rho}=c+e\ln D_{\rho} ({D_{\rho}}<{%
D_{\rho}^{E}})$, where $a$, $b$, $c$ and $e$ are fitting parameters and ${%
D_{\rho}^{E}}$ is the effective interaction width of density
transition layer. When ${D_{\rho}}>{D_{\rho}^{E}}$ the growth rate
$\gamma_{\rho}$ goes to nearly a constant and the destabilizing
effect of density gradient on KHI is fixed. These can be understood
as follows. As noted above, in the classical case, the square of the
linear growth rate is $\gamma _{c}^{2}=k^{2}\rho _{1}\rho
_{2}(v_{1}-v_{2})^{2}/(\rho _{1}+\rho _{2})^{2}\propto
(1-A^{2})\Delta v^{2}$. The wider velocity transition layer reduces
the the local velocity difference $\Delta v$ and the wider density
transition zone reduces the local Atwood number $A$. Therefore, the
linear growth rate can be decreased by ${D_{v}}$ but made larger by
$D_{\rho }$. In practical system, both the two transition layers
exist and there is a competition between these two gradient effects.
One can use the hybrid effects of density and velocity transition
layers to stabilize the KHI. By incorporating an appropriate
equation of state, or equivalently, a free energy functional, or an
external force, the present model may be used to simulate the
liquid-vapor transition and the surface tension effects on KHI.

\section*{Acknowledgements}

The authors sincerely thank the anonymous reviewers for their
valuable comments and suggestions, thank Drs. Lifeng Wang, Qing Li,
Conghai Wu, Feng Chen, Pengcheng Hao, and Yinfeng Dong for helpful
discussions. AX and GZ acknowledge support of the Science
Foundations of LCP and CAEP [under Grant Nos. 2009A0102005,
2009B0101012], National Natural Science Foundation of China [under
Grant No. 11075021]. YG and YL acknowledge support of National Basic
Research Program (973 Program) [under Grant No. 2007CB815105],
National Natural Science Foundation of China [under Grant No.
11074300], Fundamental research funds for the central university
[under Grant No. 2010YS03], Technology Support Program of LangFang
[under Grant Nos. 2010011029/30/31], and Science Foundation of NCIAE
[under Grant No. 2008-ky-13].


\begin{thebibliography}{99}
\bibitem{Succi-Book} S. Succi, \textit{The Lattice Boltzmann Equation for
Fluid Dynamics and Beyond}, (Oxford University Press, New York,
2001).

\bibitem{Yeomans} M. R. Swift, W. R. Osborn, and J. M. Yeomans, Phys. Rev.
Lett. \textbf{75}, 830 (1995); G. Gonnella, E. Orlandini, and J. M.
Yeomans, Phys. Rev. Lett. \textbf{78}, 1695 (1997); A. J. Wagner and
J. M. Yeomans, Rev. Lett. \textbf{80}, 1429 (1998); D. Marenduzzo,
E. Orlandini, and J. M. Yeomans, Phys. Rev. Lett. \textbf{92},
188301 (2004).

\bibitem{XGL1} Aiguo Xu, G. Gonnella, and A. Lamura, Phys. Rev. E \textbf{67},
056105 (2003); Phys. Rev. E \textbf{74}, 011505 (2006); Physica A \textbf{331%
}, 10 (2004); Physica A \textbf{344}, 750 (2004); Physica A
\textbf{362}, 42 (2006); Aiguo Xu, G. Gonnella, A. Lamura, G. Amati,
and F. Massaioli, Europhys. Lett. \textbf{71}, 651 (2005).

\bibitem{Sofonea-multiphase} V. Sofonea and K. Mecke, Eur. Phys. J. B \textbf{8}%
, 99 (1999); V. Sofonea, A. Lamura, G. Gonnella, and A. Cristea,
Phys. Rev. E \textbf{70}, 046702 (2004); G. Gonnella, A. Lamura, and
V. Sofonea, Phys. Rev. E \textbf{76}, 036703 (2007); A. Cristea, G.
Gonnella, A. Lamura, and V. Sofonea, Commun. Comput. Phys.
\textbf{7}, 350 (2010).

\bibitem{Chensy-PRL-1991} S. Y. Chen, H. D. Chen, D. Martinez, and W.
Matthaeus, Phys. Rev. Lett. \textbf{67}, 3776 (1991).

\bibitem{Succi-PRA-1991} S. Succi, M. Vergassola and R. Benzi, Phys. Rev. A
\textbf{43}, 4521 (1991).

\bibitem{CICP-2008} G. Vahala, B. Keating, M. Soe, J. Yepez, L. Vahala, J. Carter, and
S. Ziegeler, Comm. Comp. Phys. \textbf{4}, 624 (2008).

\bibitem{reactive-1} J. Yepez, Int. J. Mod. Phys. C \textbf{12}, 1285 (2001).

\bibitem{reactive-2} G. P. Berman, A. A. Ezhov, D. I. Kamenev, and J. Yepez,
Phys. Rev. A \textbf{66}, 012310 (2002).

\bibitem{reactive-3} K. Furtado and J. M. Yeomans, Phys. Rev. E \textbf{73},
066124 (2006).

\bibitem{Sunch-PRE-1998} C. H. Sun, Phys. Rev. E \textbf{58}, 7283 (1998).

\bibitem{Katato} T. Kataoka and M. Tsutahara, Phys. Rev. E \textbf{69},
035701(R) (2004); Phys. Rev. E \textbf{69}, 056702 (2004).

\bibitem{Watari} M. Watari and M. Tsutahara, Phys. Rev. E \textbf{67}, 036306
(2003); Phys. Rev. E \textbf{70}, 016703 (2004).

\bibitem{Xu-compressible} Aiguo Xu, Europhys. Lett. \textbf{69}, 214 (2005);
Phys. Rev. E \textbf{71}, 066706 (2005); Prog. Theor. Phys. (Suppl.) \textbf{%
162}, 197 (2006).

\bibitem{Xu-compressible-2} Yanbiao Gan, Aiguo Xu, Guangcai Zhang, Xijun Yu,
and Yingjun Li, Physica A \textbf{387}, 1721 (2008).

\bibitem{Xu-compressible-3} Feng Chen, Aiguo Xu, Guangcai Zhang, Yingjun
Li, and Sauro Succi, Europhys. Lett. \textbf{90}, 54003 (2010).

\bibitem{QuKun} K. Qu, C. Shu, and Y. T. Chew, Phys. Rev. E \textbf{75},
036706 (2007).

\bibitem{HYL} Q. Li, Y. L. He, Y. Wang, and W. Q. Tao, Phys. Rev. E \textbf{%
76}, 056705 (2007); Q. Li, Y. L. He, Y. Wang, and G. H. Tang, Phys.
Lett. A \textbf{373}, 2101 (2009); Y. Wang, Y. L. He, T. S. Zhao, G.
H. Tang, and W. Q. Tao, Int. J. Mod. Phys. C \textbf{18}, 1961
(2007).

\bibitem{PDE-1} J. Y. Zhang and G. W. Yan, Phys. Rev. E \textbf{81}, 066705
(2010); Chaos \textbf{20}, 023129 (2010).

\bibitem{kinetic nature} Q. F. Wu and W. F. Chen, \textit{DSMC method for heat
chemical nonequilibrium flow of high temperature rarefied gas},
(National Defence Science and Technology University Press, Beijing,
1999), (in Chinese).

\bibitem{KH-BOOK} S. Chandrasekhar, \textit{Hydrodynamic and Hydromagnetic
Stability}, (Oxford University, London, 1961).

\bibitem{PFA-1993} J. S. Walker,  G. Talmage, S. H. Brown, and N.
A. Sondergaard, Phys. Fluids A \textbf{5}, 1466 (1993).

\bibitem{PF-1982}  H. H. Bau, Phys. Fluids \textbf{25}, 1719 (1982).

\bibitem{PF-1980} J. W. Miles, Phys. Fluids \textbf{23}, 1719 (1980).

\bibitem{wang-pop-2009} L. F. Wang, C. Xue, W. H. Ye, and Y. J. Li, Phys.
Plasma \textbf{16}, 112104 (2009).

\bibitem{wang-pop-2010} L. F. Wang, W. H. Ye, and Y. J. Li, Phys. Plasma
\textbf{17}, 042103 (2010).

\bibitem{PRL-1999} A. Miura, Phys. Rev. Lett. \textbf{83}, 1586 (1999).

\bibitem{PRL-2002} R. Blaauwgeers, V. B. Eltsov, G. Eska, A. P. Finne, R. P.
Haley, M. Krusius, J. J. Ruohio, L. Skrbek, and G. E. Volovik, Phys.
Rev. Lett. \textbf{89}, 155301 (2002).

\bibitem{PRE-2004} G. Bodo, A. Mignone, and R. Rosner, Phys. Rev. E \textbf{%
70}, 36304 (2004).

\bibitem{POP-2005} W. Horton, J. C. Perez, T. Carter, and R. Bengtson, Phys.
Plasmas \textbf{12}, 22303 (2005).

\bibitem{wang-epl-2009} L. F. Wang, W. H. Ye, Z. F. Fan, Y. J. Li, X. T.
He, and M. Y. Yu, Europhys. Lett. \textbf{86}, 15002 (2009).

\bibitem{HEDP} R. P. Drake, \textit{High-Energy-Density Physics:
Fundamentals, Inertial Fusion and Experimental Astrophysics},
(Springer, New York, 2006).

\bibitem{turbulent} M. Lesieur, \textit{Turbulence in Fluids}, (Kluwer
Academic Publishers, Dordrecht, 1997).

\bibitem{supernovae-1} V. N. Gamexo, A. M. Khokhlo, E. S. Oran, A. Y.
Chtchelkanova, and R. O. Rosenberg, Science \textbf{299}, 77 (2003).

\bibitem{supernovae-2} A. Burrows, Nature (London) \textbf{430}, 727 (2000).

\bibitem{supernovae-3} K. Nomoto, K. Iwamoto, and N. Kishimoto, Science
\textbf{276}, 1378 (1997).

\bibitem{earth magnet} H. Hasegawa, M. Fujimoto, T. D. Phan, H. Rème, A. Balogh, M. W. Dunlop, C. Hashimoto, and R. TanDokoro,
Nature (London) \textbf{430}, 755 (2004).

\bibitem{pop-2010-2} L. F. Wang, W. H. Ye, and Y. J. Li, Phys. Plasma
\textbf{17}, 052305 (2010).

\bibitem{epl-2010-2} L. F. Wang, W. H. Ye, Z. F. Fan, and Y. J. Li, Europhys.
Lett. \textbf{90}, 15001 (2010).

\bibitem{ICF-3} B. A. Remington, R. P. Drake, H. Takabe, and D. Arnett, Phys.
Plasmas \textbf{7}, 1641 (2000).

\bibitem{ICF-4}  B. A. Remington, R. P. Drake, and D. D. Ryutov, Rev. Mod.
Phys. \textbf{78} 755 (2006).

\bibitem{ICF-book} S. Atzeni and J. Merer-ter-Vehn, \textit{The Physics of
Inertial Fusion}, (Oxford University Press, New York, 2004).

\bibitem{KH-RM-RT} G. Chimonas, Phys. Fluids \textbf{29}, 2061 (1986).

\bibitem{WENO-5th} G. S. Jiang and C. W. Shu, J. Comput. Phys. \textbf{126}, 202
(1996).

\bibitem{ENO} C. W. Shu and S. Osher, J. Comput. Phys. \textbf{77},
439 (1988).

\bibitem{WENO-4th} X. D. Liu, S. Osher, and T. Chan, J. Comput. Phys. \textbf{115}, 200 (1994).

\bibitem{SOD} G. A. Sod, J. Comput. Phys. \textbf{27}, 1 (1978).

\bibitem{NND} H. X. Zhang, Acta Aerodyn. Sin. \textbf{6}, 143 (1988).

\bibitem{outflow-KHI-1} R. Keppens and G. Tóah, Phys. Plasma \textbf{6}%
, 1461 (1999).

\bibitem{outflow-KHI-2} M. Perucho, M. Hanasz, J. M. Marti, and J. A. Miralles,
Phys. Rev. E \textbf{75}, 056312 (2007).

\bibitem{outflow-KHI-3} L. F. Wang, A. P. Teng, W. H. Ye, C. Xue, Z. F. Fan, and Y. J. Li,
Commun. Theor. Phys. \textbf{52}, 694, 056312 (2009).

\bibitem{ek} U. V. Amerstorfer, N. V. Erkaev, U. Taubenschuss, and H. K.
Biernat, Phys. Plasma \textbf{17}, 072901 (2010).

\bibitem{ek2} Ashwin J. and R. Ganesh, Phys. Rev. Lett. \textbf{104}, 215003
(2010).

\bibitem{ek3} M. Obergaulinger, M. A. Aloy, and E. M\"{u}ller, A\&A \textbf{%
515}, A30 (2010).
\end{thebibliography}
\end{document}